# How to account for quantum non-locality: ontic structural realism and the primitive ontology of quantum physics

Michael Esfeld
University of Lausanne, Department of Philosophy
Michael-Andreas.Esfeld@unil.ch

**Abstract**

The paper has two aims: (1) it sets out to show that it is well motivated to seek for an account of quantum non-locality in the framework of ontic structural realism (OSR), which integrates the notions of holism and non-separability that have been employed since the 1980s to achieve such an account. However, recent research shows that OSR on its own cannot provide such an account. Against this background, the paper argues that by applying OSR to the primitive ontology theories of quantum physics, one can accomplish that task. In particular, Bohmian mechanics offers the best prospect for doing so. (2) In general, the paper seeks to bring OSR and the primitive ontology theories of quantum physics together: on the one hand, in order to be applicable to quantum mechanics, OSR has to consider what the quantum ontology of matter distributed in space-time is. On the other hand, as regards the primitive ontology theories, OSR provides the conceptual tools for these theories to answer the question of what the ontological status of the wave-function is.

*1.    Quantum non-locality: the state of the art*

Schrödinger considered entanglement to be "*the* characteristic trait of quantum mechanics, the one that enforces its entire departure from classical lines of thought" (1935, p. 555). This assessment is confirmed by Bell's theorem (1964, reprinted in Bell 2004, ch. 2) and the subsequent experiments (e.g. Aspect et al. 1982), which are widely taken to establish that quantum entanglement gives rise to a certain sort of non-locality in nature. The argument of this paper is that structuralism in the sense of ontic structural realism (OSR) provides the best prospect to achieve an account of quantum non-locality, if it is combined with the primitive ontology theories of quantum mechanics (QM), in particular Bohmian mechanics. Furthermore, in general, the aim of this paper is to bring OSR together with the primitive ontology theories of QM, showing how they can profit from each other.

This section recalls the state of the art in the debate about quantum non-locality – in particular the role that holism, non-separability and OSR play in this debate – and explains why OSR needs to pay heed to the primitive ontology of QM. The next section briefly introduces the three worked out primitive ontology theories of QM – namely Bohmian mechanics, the GRW matter density theory and the GRW flash theory – and shows how, by making use of OSR, these theories can achieve a complete and coherent account of the ontological status of the wave-function. Section 3 then investigates whether and how one can accomplish a satisfactory explanation of quantum non-locality by bringing OSR together with these theories.

Consider the EPR experiment (Einstein, Podolsky and Rosen 1935, Bohm 1951, pp. 611-622): two elementary particles are emitted together from a source and then move apart in opposite directions. When the two particles are separated by a spacelike interval, parameters to be measured on each of them are fixed in each wing of the experiment by Alice and Bob respectively, and measurement outcomes are obtained. Bell's theorem establishes that the objective probabilities for the measurement outcome in one wing of the experiment depend on what happens in the other wing, although both wings are separated by a spacelike interval; there is no common cause in the common past of both wings possible that could screen the probabilities for what happens in the one wing off from what happens in the other wing.

More precisely, let us denote by $\lambda$ whatever in the past may influence the behaviour of the measured quantum systems according to the theory under consideration (which may be textbook QM, or a theory that admits additional, so-called hidden variables). In particular, the state of the two particle system at the source of the experiment is part of $\lambda$. Let $a$ stand for Alice's measurement setting in her wing of the experiment and $A$ for Alice's outcome. Furthermore, let $b$ stand for Bob's measurement setting in his wing of the experiment and $B$ for Bob's outcome. Bell's principle of locality can then be formulated in the following manner:

$$P_{a,b}(A|B,\lambda) = P_a(A|\lambda) \tag{1}$$

$$P_{a,b}(B|A,\lambda) = P_a(B|\lambda)$$

That is to say: the probabilities for Alice's outcome are fixed by conditionalizing only on her measurement setting and $\lambda$. Conditionalizing also on Bob's setting and outcome does not change the probabilities for Alice's outcome. Thus, the probabilities for Alice's outcome do not depend on what Bob's setting and outcome may be. The same goes for Bob. The following figure illustrates this situation:

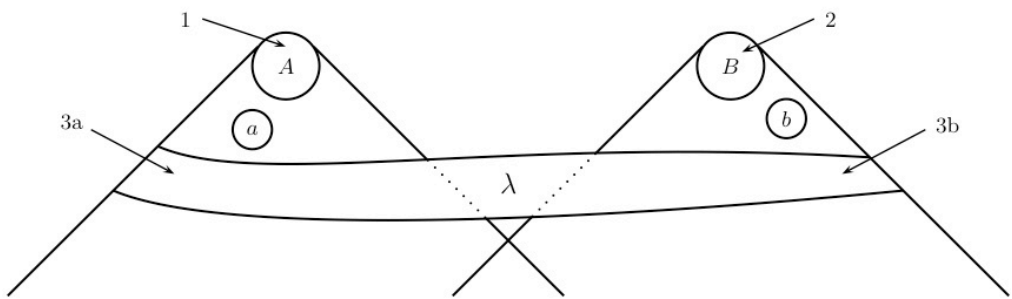

*Figure 1: The situation that Bell considers in the proof of his theorem. Figure taken from Seevinck (2010, appendix) with permission of the author.*

Bell's theorem then proves that QM violates (1). Moreover, any theory that reproduces the well-confirmed experimental predictions of QM has to violate (1) (see Bell 2004, ch. 24). One can therefore say that Bell's theorem puts a constraint on any – present or future – physical theory that is to match the experimentally confirmed predictions of QM. Violating (1) means that conditionalizing on the setting and the outcome in the one wing of the experiment changes the probabilities for the outcome to be obtained in the other wing, although both wings are separated by a spacelike interval. That is why the violation of (1) is widely taken to establish non-locality in the following sense: in some situations, what happens at a given space-time point is influenced not only by what there is in the past light-cone of that event, but also by what happens at points that are separated from the point in question by a spacelike interval. Thus, Maudlin (2011, chs. 1-6) speaks of non-local influences in his seminal monograph on Bell's theorem (see in particular pp. 118-119, 135-141), and this analysis is strengthened by recent work that seeks for precise formulations of Bell's locality principle.[1]

However, it is possible to avoid the conclusion of non-locality by taking the quantum probabilities only as a guide for the rational credences of agents given the evidence that is available to them instead of objective probabilities, as argued notably by Fuchs (2010) and Healey (2012). The price for this stance then is that Bell's theorem – and quantum mechanics in general – does not tell us anything about the way the world is. As Healey (2012, p. 730) puts it, "quantum states convey knowledge or information concerning a system or ensemble without describing its physical condition". Note that, *pace* Healey (2012, section 4.5, in particular p. 759), recent EPR-type experiments do not provide evidence against a realist understanding of quantum entanglement in terms of a physical relation and its consequence, namely the mentioned non-locality (see Egg 2013).

The present paper accepts quantum non-locality as its starting point. The task then is to develop an explanation of quantum non-locality. Explaining non-locality, however, does not have to mean answering the question of why nature is non-local. Non-locality, if it obtains, is a fundamental feature of nature. Thus, one can maintain that there can be no explanation of why nature is non-local, in the same way as there can be no explanation of why nature is local, if it were local (as was assumed in classical field theories). Nonetheless, what can with reason be called for is an account of non-locality in the sense of putting forward an ontology that accommodates Bell's theorem and the empirical evidence for quantum non-locality.

One can roughly distinguish four main types of attempts to achieve such an ontology in the literature:

a) One may envisage a direct causal account in the sense of admitting a signal, force or field that transmits information with superluminal velocity from the one wing of the EPR experiment to the other wing. However, no such account has been elaborated on in the literature.[2] Simply adding to classical physics a new type of signal, force or field that violates classical field theory by propagating with superluminal velocity would clearly be an *ad hoc* move. Moreover, classical forces or fields have properties of the particles as their source (such as mass and charge), whereas there are no properties of the particles that could be the source of such a quantum force. Finally, recent experiments put a substantial lower bound on the velocity with which such a signal, force or field would have to propagate. That lower bound is likely to increase further as better measurement techniques become available.[3]

---

[1] See notably Norsen (2009), Seevinck (2010) and Seevinck and Uffink (2011) as well as the encyclopedia entries on Bell's theorem by Shimony (2009) and by Goldstein et al. (2011).

[2] Chang and Cartwright (1993, section III) consider such an account, but do not work it out.

[3] See Salart et al. (2008) and Cocciaro, Faetti and Fronzoni (2011 and 2013).

b) One may search for a loophole in the proof of Bell's theorem that enables one to maintain that the non-locality following from the violation of (1) is only apparent: the proof of this theorem requires not only the mentioned principle of locality (1), but also the assumption that the measurement settings *a* and *b* are independent of λ. Hence, it is in principle possible to explain the EPR experiment by rejecting that independence assumption, while retaining locality (1). Failure of such independence can arise in two ways: either λ influences the measurement settings, or the measurement settings exert some influence on λ. In the former case, one can postulate a local common cause of the measurement outcomes in the intersection of their past light-cones that accounts for the settings *a* and *b* as well as the outcomes *A* and *B*.[4] In the latter case, one has to postulate influences travelling backwards in time. Thus, either the measurement settings *a* or *b*, or the outcomes *A* or *B*, which depend on the settings, would have to change λ retroactively.[5]

However that may be, the general problem with rejecting the mentioned independence assumption is that it is not specific for Bell's theorem. The presupposition of the measurement settings being independent of the prior state of the measured system applies to any experimental evidence. Furthermore, this assumption does not imply any sort of indeterminism, or the free will of experimental scientists. It can be satisfied in a deterministic theory by assuming that what determines λ and what determines *a* and *b* is distinct in the initial state of the universe (see Bell et al. 1985).

c) The non-locality that is established by Bell's theorem and the subsequent experiments concerns the behaviour of matter in physical space. Hence, if one abandons the presupposition that the physical reality is situated in four-dimensional space-time, one can avoid a problem with locality. Thus, the stance that is known as wave-function realism maintains that the universal wave-function, which is situated in a very high-dimensional space, is the physical reality.[6] Consequently, as long as the dynamics of the wave-function is given by a linear and deterministic equation, no violation of any locality principle occurs in that high-dimensional space.

However, in the first place, one faces the problem how to account for our experience within an ontology of a wave-function existing in a very high-dimensional space. Moreover, one can object that it would be well-motivated only as a last resort to abandon the assumption that the physical reality consists in matter distributed in three-dimensional space or four-dimensional space-time and that the task of mechanics is to explain the temporal development of that distribution of matter (see Monton 2006). But at least as far as Bell's theorem is concerned, this theorem tells us something about a physical reality in four-dimensional space-time, namely that events that are separated from a given space-time point by a spacelike interval can in certain situations nevertheless influence what happens at that point.

d) One may claim that QM, by contrast to classical mechanics, involves some sort of holism, non-separability or structuralism, which is expressed in the quantum formalism by the entanglement of the wave-function of two or more quantum systems. In brief, the idea is this one: due to the entanglement of the wave-function, quantum systems are not separate individuals, but their temporal development occurs in tandem so to speak, whatever their spatial distance may be.

---

[4] See San Pedro (2012) and Hofer-Szabó, Rédei and Szabó (2013, ch. 9) for an overview and discussion.

[5] See notably Price (1996, ch. 8 and 9). See furthermore the papers in *Studies in History and Philosophy of Modern Physics* 38 (2008), pp. 705-784.

[6] See notably Albert (1996) and the papers in Albert and Ney (2013) for discussion. That space cannot be considered as the configuration space of the universe in this case, since there is no configuration of anything in another space (i.e. three-dimensional space or four-dimensional space-time) that the points of this high-dimensional space represent.

If one is not inclined to go into (b) or (c) for the general reasons indicated above, one has, as things stand, the following choice for a quantum ontology that accommodates Bell's theorem: one can either seek to amend classical mechanics in terms of a signal, force or field that transmits information with superluminal velocity (a) or one has to develop an account of the temporal development of matter that breaks with the paradigm set by classical mechanics (d). Given that the former is *ad hoc*, there is a good reason to try out the latter.

Indeed, since the 1980s, the entanglement of the wave-function of a two-particle-system such as the one in the EPR experiment has been conceived in terms of non-separability and holism (e.g. Howard 1985, 1989, Teller 1986, Healey 1991). Later, this conception has been elaborated on in terms of ontic structural realism (OSR) (Ladyman 1998, French and Ladyman 2003). OSR provides a precise expression of the idea of non-separability (see Esfeld 2004): the two-particle-system in the EPR experiment (and, in general, any entangled quantum system) instantiates a structure defined over a domain of two or more quantum objects. That structure consists in a concrete physical relation of entanglement relating all the objects in the domain in question. That relation is such that (a) it does not supervene on intrinsic properties of the objects in question and (b), moreover, it does not even require objects with an intrinsic identity as its relata. This structure captures non-separability because due to the relation of entanglement, the temporal development of the objects in the domain of the structure in question is tied together. This structure therefore is the reason why the measurement outcomes in the EPR experiment are correlated as predicted by the quantum formalism. Furthermore, the proponents of OSR conceive the structures to which they are committed as modal, grounding laws of nature such as the laws of QM (see e.g. Ladyman and Ross 2007, chs. 2-5, French 2014, chs. 9-11), or even as being causal, bringing the measurement outcomes about (see Esfeld 2009).

For a long time, it was thought that invoking non-separability, holism or OSR is on its own sufficient to achieve a satisfactory account of the EPR experiment. However, recent research has made clear that non-separability or OSR neither avoids the commitment to non-locality nor offers as such an account of non-locality (see notably Esfeld 2013 and Henson 2013). The reason is, in brief, that non-separability and OSR tell us that the temporal development of quantum systems that are related by a relation of entanglement is tied together, but on their own, they do not include an account of how space-like separated, correlated measurement outcomes are obtained in the EPR experiment.

Nonetheless, the claim of this paper is that (1) there is something new in quantum non-locality, which, as Schrödinger put it in the quotation at the beginning of this paper, "enforces its entire departure from classical lines of thought" and that (2) invoking holism, non-separability or structuralism in the sense of OSR is on the right track to capture that new feature, although on its own insufficient to accomplish an account of quantum non-locality. What is needed is to employ the guiding idea of OSR to develop an account of the dynamics of quantum systems in space-time such that the correlated outcomes of the EPR experiment become intelligible. In order to work out such an account, it is obvious that one has to consider what exactly the distribution of matter in physical space is. In other words, one has to turn to what is known as the primitive ontology of QM.

Generally speaking, there is a cogent reason for the proponents of OSR to take up the primitive ontology theories of QM: OSR cannot adopt the stance according to which the physical reality consists in the universal wave-function in a high-dimensional space (see (c) above). One could envisage maintaining OSR with respect to the mathematical properties of that space, as one can put forward OSR with respect to the metric of the four-dimensional space-time of general relativity theory (see e.g. Esfeld and Lam 2008). However, as regards matter, if the universal wave-function were the physical reality, there would be only one object, which, moreover, would develop according to a local dynamics in that high-dimensional space. Alternatively, since the wave-function is a field on that space, one could also go for the stance that it consists in a plurality of pointlike objects, namely the field values instantiated at the points of that space. But these are in any case intrinsic properties, so that this stance results in the atomistic view of everything supervening on the distribution of intrinsic properties instantiated at the points of a very high-dimensional space (see Albert 1996, p. 283, note 7). Hence, if the wave-function were the physical reality, there would be nothing in quantum physics for OSR to hook up to. Thus, in order to be applicable to QM, OSR has to regard the formalism of QM as referring to a physical reality of matter distributed in ordinary physical space that is entangled and whose dynamics therefore is non-local. In brief, in order to be applicable to QM, OSR has to consider the primitive ontology of QM.

Accordingly, the aim of this paper is to apply OSR to the primitive ontology theories of QM in order to achieve the account of quantum non-locality that structuralism cannot provide on its own. Or, to put it differently, the aim of this paper is to make use of the conceptual tools of structuralism in order to examine whether and how the primitive ontology theories of QM can set out a satisfactory account of quantum non-locality.

## 2.     *The primitive ontology of quantum mechanics*

The primitive ontology theories hold that QM is about matter distributed in three-dimensional space or four-dimensional space-time and its temporal development, by contrast to being about the universal wave-function in a very high-dimensional space.[7] The ontology of matter distributed in physical space is primitive in the sense that it cannot be inferred from the formalism of textbook QM, but has to be put in as the referent of that formalism. The motivation for doing so is to obtain an ontology that can account for the existence of measurement outcomes – and, in general, the existence of the macrophysical objects with which we are familiar before doing science. Hence, what is introduced as the primitive ontology has to be such that it can constitute measurement outcomes and localized macrophysical objects in general. That is why the primitive ontology consists in one actual distribution of matter in space at any time (no superpositions), and the elements of the primitive ontology are localized in space-time, being "local beables" in the sense of Bell (2004, ch. 7), that is, something that has a precise localization in space at a given time.

There are three elaborate primitive ontology theories of QM. The de Broglie-Bohm theory, going back to de Broglie (1928) and Bohm (1952) and known today as Bohmian mechanics (BM) (see Dürr, Goldstein and Zanghì 2013) is the oldest of them. BM endorses particles as the primitive ontology, maintaining that there is at any time one actual configuration of particles in three-dimensional space, with the particles moving on continuous trajectories in space. BM therefore needs two laws: the guiding equation fixing the temporal development of the position of the particles, and the Schrödinger equation determining the temporal development of the wave-function. These two laws are linked in the following manner: the role of the wave-function, developing according to the Schrödinger equation, is to determine the velocity of each particle at any time $t$ given the position of all the particles at $t$.

---

[7]   The term "primitive ontology" goes back to Dürr, Goldstein and Zanghì (2013, ch. 2, see end of section 2.2, paper originally published 1992).

Furthermore, there are two primitive ontology theories of QM using the dynamics proposed by Ghirardi, Rimini and Weber (GRW) (1986), which seeks to include the textbooks' postulate of the collapse of the wave-function upon measurement into a modified Schrödinger equation. Ghirardi, Grassi and Benatti (1995) develop an ontology of a continuous matter density distribution in physical space (GRWm): the wave-function in configuration space and its temporal development as described by the GRW equation represent at any time the density of matter in physical space. The spontaneous localization of the wave-function in configuration space (its "collapse") represents a spontaneous contraction of the matter density in physical space, thus accounting for measurement outcomes and well localized macrophysical objects in general (see also Monton 2004).

The other theory goes back to Bell (2004, ch. 22, originally published 1987): whenever there is a spontaneous localization of the wave-function in configuration space, that development of the wave-function in configuration space represents an event occurring at a point in physical space. These point-events are today known as flashes; that term was introduced by Tumulka (2006). According to the GRW flash theory (GRWf), the flashes are all there is in space-time. Hence, the temporal development of the wave-function in configuration space does not represent the distribution of matter in physical space. It represents the objective probabilities for the occurrence of further flashes, given an initial configuration of flashes. There thus is no continuous distribution of matter in physical space, namely no trajectories or worldlines of particles, and no field – such as a matter density field – either. There only is a sparse distribution of single events in space-time. Although GRWf and GRWm are rival proposals for an ontology of the same formalism (the GRW quantum theory), there also is a difference between them on the level of the formalism: if one endorses the GRWm ontology, it is reasonable to pursue a formalism of a continuous spontaneous localization of the wave-function (as done in Ghirardi, Pearle and Rimini 1990), whereas if one subscribes to the GRWf ontology, there is no point in doing so.

These theories hence put forward different proposals about the nature of matter, which cover the main metaphysical conceptions of matter – particles or atoms, stuff or gunk, single events. Nonetheless, their structure is the same: they consist in a proposal for a primitive ontology of matter distributed in physical space and a law for its temporal development (see Allori et al. 2008). What is the status of the wave-function in these theories? More precisely, what is the status of the universal wave-function, that is, the wave-function of the whole configuration of matter in the universe?

On the one hand, it would be incoherent to regard the wave-function as a physical entity on a par with the primitive ontology: it could not be a field in physical space, because it does not assign values to points of physical space. If it were a field, it could only be a field in the high-dimensional mathematical space each point of which represents a possible configuration of matter in physical space. However, this relationship of representation could not be turned into a causal relation: it would be mysterious how an entity existing in configuration space could exert a physical influence on the temporal development of entities existing in physical space. Against this background, it is well-motivated to regard the universal wave-function as nomological, namely as being part of the law of the development of the primitive ontology, by contrast to being a physical entity on a par with the primitive ontology (see Dürr, Goldstein and Zanghì 2013, ch. 12, in the context of BM).

However, on the other hand, simply claiming that the universal wave-function is nomological clearly is an incomplete answer to the question of what the status of the wave-function is. In the first place, even the universal wave-function develops itself in time according to a law, namely the Schrödinger equation – unless one takes for granted that the Schrödinger equation will eventually be replaced by the Wheeler-deWitt equation, in which a stationary wave-function figures. But even if the universal wave-function turned out to be stationary, there would still remain the fact that there are many different universal wave-functions possible all of which fit into the same law (the Schrödinger equation, or the Wheeler-deWitt equation). Consequently, there is a good reason to regard the wave-function not only as nomological, but also as encoding a concrete physical fact over and above the primitive ontology: the latter is not sufficient to determine the (universal) wave-function. There can be two or more identical (universal) configurations of particles, matter density or flashes and yet different wave-functions of these configurations, leading to different temporal developments of them. In other words, the universal wave-function also has the feature of an intitial condition: it and the initial configuration of matter have to be fed into the law of motion (e.g. the guiding equation in BM) in order to obtain as a result the temporal development of the configuration of matter.

Adopting Bell's terminology, one may be inclined to simply say that the universal wave-function represents a "non-local beable" that exists over and above the "local beables" making up the primitive ontology. But such a proposition would simply state the issue instead of resolving it as long as one does not elaborate on what that non-local beable is. Here is the meeting point of primitive ontology theories of QM and OSR: if one adopts OSR, one can spell out what the "non-local beable" represented by the universal wave-function is and thereby provide a complete and coherent answer to the question of what the status of the wave-function in these theories is. That is to say, the wave-function represents a structure that is instantiated by the elements of the primitive ontology – the particles, the matter density values, or the flashes at points in physical space. That structure is a concrete physical entity in space-time in the sense that it consists in a network of relations of entanglement encompassing all these elements. Thus, in BM, that structure consists in a network of relations that relates the positions of strictly speaking all the particles with each other such that the temporal development of the position of each particle (i.e. its velocity) depends on the positions of all the other particles. Nonetheless, that structure exists over and above the elements of the primitive ontology, since it does not supervene on whatever intrinsic features these elements may have. In particular, it neither supervenes on their position in space at any time, nor on their spatio-temporal relations (nor on intrinsic properties such as the mass and the charge of these elements). In being a concrete physical entity, that structure can develop in time.

The decisive issue then is that this structure is modal in the same sense as intrinsic properties of the particles in classical mechanics – such as mass and charge – are modal: the distribution of these properties in physical space at any given time grounds force laws that describe the acceleration of the particles (Newton's law of gravitation, Coulomb's law) in the sense that these laws supervene on the distribution of these properties (given the distribution of these properties in space and a constant, these laws are fixed). By the same token, the structure represented by the quantum mechanical wave-function, taking all the elements of the primitive ontology as its relata, grounds the law for the temporal development of these elements in the sense that this law supervenes on that structure (i.e. the Bohmian guiding equation or the GRW equation). As the same force laws can supervene on different distributions of mass and charge in space, so the same law can supervene on different such structures, represented by different universal wave-functions. One can then further elaborate on the modal nature of that structure by taking that structure to be a nomological fact instantiated in the world (cf. the primitivism about laws applied to QM in Maudlin 2007), or by regarding it as a holistic and dispositional property that relates the elements of the primitive ontology and that manifests itself in a certain temporal development of these elements.[8]

The decisive point is that the primitive ontology theories of QM on the one hand and OSR on the other can help each other. For the primitive ontology theories of QM, OSR provides a complete and coherent answer to the question of what the ontological status of the universal wave-function in these theories is: OSR yields a precise physical sense for the notion of a "non-local beable" represented by the wave-function in terms of a structure that is instantiated by and that relates all the elements of the primitive ontology, tying their temporal development together. For OSR, the primitive ontology theories of QM provide concrete physical entities in physical space that instantiate the structure in question, given the fact that OSR is applicable to QM only if it commits itself to a distribution of matter in physical space as the referent of the formalism of QM (see the end of the preceding section). On this basis, let us now investigate how the marriage of OSR and the primitive ontology theories of QM can achieve the account of quantum non-locality that OSR cannot accomplish on its own, as argued in the preceding section.

3. *Primitive ontology, OSR and the account of quantum non-locality*

To start with, consider what BM says about the EPR experiment. Since BM is a deterministic theory, it has to reject the condition known as parameter independence; on a deterministic theory, it is determined what the outcome of a measurement is before the measurement actually occurs. Thus, according to BM, the two measurement outcomes of the EPR experiment are determined by what there is in the past of the two wings of the experiment (represented by $\lambda$ in figure 1 in section 1), the parameters that are measured in both wings of the experiment (represented by *a* and *b* in figure 1) and the positions of the particles that constitute the measurement apparatuses in both wings. Hence, following BM, fixing the parameter in one wing of the EPR experiment influences the trajectory of the particles in *both* wings. However, there is no direct interaction among the particles: no particle acts on another particle in BM. Fixing the parameter in one wing of this experiment influences the trajectory of the particles in both wings via the wave-function of the whole system, which has the job to determine the velocity of each particle at *t*, given the position of all the particles at *t*.[9] That notwithstanding, BM supports counterfactuals of the following type: "If Alice had chosen a different setting in her wing of the EPR experiment, Bob would have obtained a different outcome in his wing of the experiment". Nonetheless, there is no direct influence from Alice's setting to Bob's outcome. Alice's setting influences Bob's outcome only via the wave-function, which in turn influences *both* outcomes. Strictly speaking, on BM, the velocity of any particle at *t* depends on the position of *all* the other particles at *t* – including the particles making up the apparatuses – via the wave-function.

---

[8] See Belot (2012, pp. 77-80) and Esfeld et al. (2014, sections 4-5) for BM as well as Dorato and Esfeld (2010) for GRW.

Applying OSR to the wave-function in BM turns this treatment of the EPR experiment into a clear and coherent account of quantum non-locality. There is a physical structure of entanglement that takes all the particles as its relata, being instantiated by the configuration of the particles and being represented by their wave-function. By contrast to an account of quantum non-locality in terms of superluminal interaction (proposal of type (a) in section 1), there is no action at a distance among anything in BM, simply because a modal structure instantiated by the configuration of matter is another conception of the determination of the temporal development of physical objects than direct interaction among the parts of that configuration. Hence, there is quantum non-locality because the temporal development of quantum systems is not determined by properties that are intrinsic to each object (as mass and charge are intrinsic properties of particles in classical mechanics and determine their temporal development), but by a structure that is instantiated by the configuration of all the quantum objects. This explanation may sound unfamiliar given our familiarity with a classical domain of objects being moved by local forces, but it is complete and coherent. As Solé (2013) argues in his examination of Bohm's quantum theory, there is no need for a formulation in terms of a second order theory that is committed to a dubious quantum force for BM to be explanatory.

As pointed out in section 1, there can be no explanation of why nature is non-local, if it is non-local, as there can be no explanation of why nature is local, if it is local. In the latter case, there are local forces going back to intrinsic properties of the objects in nature; in the former case, there is a structure instantiated by the configuration of the objects. As there can be no further explanation of how intrinsic properties of particles such as their mass and charge give rise to forces such as gravitation and electromagnetism that determine the temporal development of the velocity of the particles, so there can be no further explanation of how a structure instantiated by the configuration of the particles determines the temporal development of the position of the particles. BM thus is a paradigm example of how a structure that takes all the physical objects as its relata can determine their temporal development, by contrast to a conception according to which the temporal development of the physical objects is determined by direct interaction among them (which would imply superluminal interaction in the EPR experiment).

Let us enquire now how GRWm handles the EPR experiment. One can also in GRWm conceive the universal wave-function as representing a structure that is instantiated by the matter density as a whole. That structure is modal in the sense that it grounds the law for the temporal development of the matter density (i.e. the GRW equation), which is a probabilistic law in this case. One can therefore say that this structure is modal in the sense that it consists in a certain propensity of the matter density as a whole that grounds the GRW probabilities and that manifests itself in a certain temporal development of the matter density, as represented by the spontaneous localization of the wave-function in configuration space (see Dorato and Esfeld 2010). However, this on its own is not sufficient as an account of quantum non-locality. The decisive issue is how the manifestation of this propensity comes about so that the correlated outcomes of the EPR experiment are explained. On GRWm, the measurement in one wing of the EPR experiment triggers this propensity in triggering a change in the shape of the matter density in both wings of the experiment, so that, in Bohm's (1951) version of the experiment, the shape of the matter density constitutes two spin measurement outcomes. By contrast to the treatment of the EPR experiment in BM, this is not a pure account of a structure determining the temporal development of its relata, but it involves features of an action at a distance account.

---

[9] See Bell (2004, ch. 4) and Norsen (2013) as to how BM accounts for the outcomes of spin measurements – such as the EPR experiment in the version of Bohm (1951, pp. 611-622) – in terms of the temporal development of the position of particles as described by the wave-function.

In order to bring out the contrast between BM and GRWm in that respect, let us turn to another thought experiment proposed by Einstein, namely the one of what is known as Einstein's boxes presented before the EPR experiment (1935) at the Solvay conference in 1927 (see Norsen 2005). Following the version of this thought experiment set out in de Broglie (1964, pp. 28-29), consider a box which is prepared in such a way that there is a single elementary quantum particle in it. The box is split in two halves which are sent in opposite directions, say from Brussels to Paris and Tokyo. When the half-box arriving in Tokyo is opened and found to be empty, there is on textbook QM as well as on all the proposals for a primitive ontology of QM a fact that the particle is in the half-box in Paris. If one takes the textbooks' postulate of the collapse of the wave-function upon measurement literally, this means that the act of opening the box in Tokyo creates the fact that there is a particle in the box in Paris. This is an illustration of what Einstein regarded as "spooky action at a distance". Since such action at a distance is absurd, Einstein maintained that the particle always travels in one of the two half-boxes, depending on its initial position, and that its motion is not influenced by whatever operation is carried out on the other half-box.

Whereas no such local account is possible in the EPR experiment, as shown by Bell's theorem, it is possible in this case and provided by BM: according to BM, there always is one particle moving on a continuous trajectory in one of the two half-boxes, and opening one of them only reveals where the particle was all the time. Following BM, the particle in the box has what is known as an effective wave-function of its own so that the position of distant particles is *de facto* irrelevant for its trajectory. On GRWm, by contrast, the particle is in fact a matter density field that stretches over the whole box and that is split in two halves of equal density when the box is split, with the two halves travelling in opposite directions. Upon interaction with a measurement device, one of these matter densities (the one in Tokyo in the example given above) vanishes, while the matter density in the other half-box (the one in Paris) increases so that the whole matter is concentrated in one of the half-boxes. This does not mean that some matter travels from Tokyo to Paris, since it is impossible to assign any speed to this change of location. For lack of a better expression let us say that some matter is delocated from Tokyo to Paris; for even if the spontaneous localization of the wave-function in configuration space is a continuous process, the time it takes for the matter density to disappear in one place and to reappear in another place does not depend on the distance between the two places.

The decisive point is that one and the same matter can disappear in Tokyo and reappear in Paris without travelling from Tokyo to Paris. Strictly speaking, this is not action at a distance, since the action is local in Tokyo; but the process of the delocation of matter is rather mysterious, or "spooky". The contrast with BM thus is that on BM, particles always move on continuous trajectories that are determined by a structure instantiated by the particle configuration, without anything ever being delocated across space. In the case of GRWm, on the contrary, the structure instantiated by the matter density as a whole is a propensity whose manifestation implies that some matter disappears in location $A$ and reappears in location $B$, without travelling from $A$ to $B$ with any speed.

As regards GRWf, the ontology of this theory can also be construed in terms of a structure that is instantiated by a configuration of flashes, that is represented by the wave-function of that configuration and that is a propensity whose manifestation consists in the occurrence of new flashes. However, since there is only a sparse distribution of flashes in space-time, a measurement operation often has to trigger the propensity of a *past* configuration of flashes. Thus, according to GRWf, in the EPR experiment, there are only two flashes at the source and no quantum system whatsoever (no particle, no field, no matter density) in either wing of the experiment. The measurement in one wing triggers the propensity of the two flashes at the source to bring about the occurrence of two new flashes, one in each wing of the experiment, such that the two flashes make up for two correlated spin measurement outcomes. In the case of the particle in the box, there only is one flash in the box at the source in Brussels, nothing travels in either box, but opening the box in Tokyo triggers the occurrence of a new flash in Paris.

Again, there is a clear contrast with BM: in BM, there is a structure instantiated by a configuration of objects with the structure instantiated at a time $t$ determining the development of the objects at that very $t$, without there being any jumps in space or time. In GRWf, by contrast, there are not only jumps in space (as in GRWm), but also jumps in time, since what is triggered is a propensity instantiated by *past* flashes, which can, moreover, amount to bringing about one single flash at an arbitrarily large distance in space (as in the case of Einstein's boxes). It is therefore difficult to see how the inference could be avoided that this is a case of action at a distance. To sum up, although all the primitive ontology theories of QM can be construed in terms of the wave-function representing a structure that is instantiated by the configuration of matter, the decisive issue is how that structure determines the temporal development of matter. There are remarkable differences among these theories as to how they spell out that determination. Only one of them, namely BM, provides a neat account of a structure determining the temporal development of the matter that this structure relates and thereby explaining non-locality, by contrast to any superluminal action or delocation of something.

In conclusion, the following is a well-grounded proposition: (a) quantum physics conceives the temporal development of matter in terms of non-separability and thus a certain sort of holism; (b) this sort of holism is made precise by ontic structural realism (OSR); (c) this conception constitutes an alternative to the way in which the temporal development of matter is construed in classical physics as well as provides the key to avoid the pitfall of what Einstein called "spooky action at a distance". However, ironically enough, although the proposals invoking non-separability, holism or OSR are usually hostile to Bohmian mechanics (BM), it is BM – and as things stand only BM – that yields the ontology of matter distributed in physical space which enables OSR to achieve the intended result, namely a clear, complete and coherent account of quantum non-locality that avoids a committed to superluminal action.